\documentclass[a4paper]{jpconf}
\usepackage{graphicx}
\usepackage{amsmath,amssymb,amsfonts}
\usepackage{subfigure}
\usepackage{listings}

\newcommand{\secdec}{{\textsc{SecDec}}}
\def\be{\begin{equation}}
\def\ee{\end{equation}}
\newcommand{\bea}{\begin{eqnarray}}
\newcommand{\eea}{\end{eqnarray}\noindent}
\newcommand{\nn}{\nonumber}
\newcommand{\bcen}{\begin{center}}
\newcommand{\ecen}{\end{center}}
\newcommand{\rd}{{\mathrm{d}}}

\def\url#1{\texttt{#1}}
\def\eps{\epsilon}

\begin{document}
\title{\textsc{SecDec}: A tool for numerical multi-loop calculations\footnote{Based on a talk given at the 14th International Workshop on Advanced Computing and
Analysis Techniques in Physics Research (ACAT), Uxbridge, London, UK, September 2011.}}

\author{S. Borowka$^a$, J. Carter$^b$ and G. Heinrich$^a$}
\address{$^a$Max-Planck Institute for Physics, Munich, Germany}
\address{$^b$Institute for Particle Physics Phenomenology, University of Durham, UK}

\ead{j.p.carter@durham.ac.uk, sborowka@mpp.mpg.de, gudrun@mpp.mpg.de}

\begin{abstract}
The version 2.0 of the program \secdec{} is described, which can be used for the 
extraction of poles within dimensional regularisation
from multi-loop integrals as well as phase space integrals.
The numerical evaluation of the resulting finite functions 
is also done by the program in an automated way, with no restriction 
on the kinematics in the case of loop integrals.
\end{abstract}

\section{Introduction}

Nowadays we are in the long awaited situation of being confronted with a 
wealth of high energy collider physics data, enabling us to explore 
physics at the TeV scale.
However, for the analysis and interpretation of these data, precise 
theory predictions are mandatory. In most cases, this means that calculations 
beyond the leading order in perturbation theory are necessary. 
Such calculations either involve integrations over loop momenta for the virtual
corrections, or integrations over phase spaces for the real radiation corrections.
In both cases, multi-dimensional parameter integrals need to be evaluated, 
which can contain ultraviolet singularities
and, in the presence of massless particles, contain soft and/or collinear 
singularities. These singularities, if regulated by dimensional regularisation, 
appear as poles in $1/\eps$, but factorising the poles from 
complicated multi-loop or multi-parameter integrals is a highly non-trivial task.
The program \secdec{}, presented in \cite{Carter:2010hi,Borowka:2012yc}, performs this task 
in an automated way, based on the algorithm of 
sector decomposition \cite{Binoth:2000ps,Roth:1996pd,Hepp:1966eg}.
Other implementations of sector decomposition in public programs can be 
found in \cite{Smirnov:2009pb,Bogner:2007cr,Gluza:2010rn}. 
However, the latter programs, including \secdec{}\,1.0\,\cite{Carter:2010hi}, 
are not designed  to cope with kinematics where physical thresholds 
can be present in the integration region.
The program \secdec{}\,2.0\,\cite{Borowka:2012yc} is able to achieve this task, 
by an automated deformation of the 
integration contour into the complex plane\,\cite{Soper:1999xk,Nagy:2006xy,Binoth:2005ff,Anastasiou:2007qb}.
In this talk the emphasis is on the presentation of the new features of the 
program \secdec.

\section{The algorithm}

\subsection{General framework}
The decomposition algorithm is described in detail in~\cite{Heinrich:2008si}, 
here we only describe the basic concepts. 
Consider an $N$-dimensional parameter integral, for example 
\bea
I_N=\int_0^1 dx_1\ldots \int_0^1 dx_N \,x_1^{-1-\eps} \,\frac{g_3(\vec{x})}{
\left[x_1\,g_1(\vec{x})+x_2\,g_2(\vec{x})\right]} \;,
\label{eq:overlap}
\eea
where $g_1,g_2$ are polynomial functions which do not vanish for $x_1,x_2\to 0$.
However, we would like to factorize the integrand such that the term in square brackets 
is non-vanishing in the limit $x_1\to 0,x_2\to 0$.
This can be achieved by a decomposition into sectors where $x_1$ and $x_2$ 
are ordered: we multiply equation~(\ref{eq:overlap}) with 
$1=\Theta(x_1-x_2)+\Theta(x_2-x_1)$ and substitute $x_2\to x_1\,x_2$ in the first sector 
and $x_1\to x_2\,x_1$ in the second sector, to arrive at
\bea
I_N=\int_0^1 dx_1\ldots \int_0^1 dx_N \,x_1^{-1-\eps} \,\left\{\frac{\tilde{g}_3}{
\left[ \tilde{g}_1(\vec{x})+x_2\,\tilde{g}_2(\vec{x})\right]}+
x_2^{-1-\eps} \,\frac{\bar{g}_3}{
\left[ x_1\,\bar{g}_1(\vec{x})+\bar{g}_2(\vec{x})\right]} \right\}\;,
\label{eq:nonoverlap}
\eea
where $\tilde{g}_i,\bar{g}_i$ are functions of the transformed variables.
As $g_1,g_2$ are non-vanishing at the origin, the terms in square brackets 
cannot lead to singularities anymore; the singularities are factored out into 
the monomials of type $x_i^{-1-\eps}$ instead, and 
subtraction of the poles and expansion in $\epsilon$ are straightforward in this form. 
In more complicated cases, the decomposition procedure may have to be iterated to achieve 
a full factorisation. A detailed description can be found e.g. 
in \cite{Binoth:2000ps,Heinrich:2008si}.

\subsection{Loop integrals}

A scalar Feynman integral $G$ in $D$ dimensions 
at $L$ loops with  $N$ propagators, where 
the propagators can have arbitrary, not necessarily integer powers $\nu_j$,  
has the following representation in momentum space:
\begin{eqnarray}\label{eq0}
G&=&\int\prod\limits_{l=1}^{L} \rd^D\kappa_l\;
\frac{1}
{\prod\limits_{j=1}^{N} P_{j}^{\nu_j}(\{k\},\{p\},m_j^2)}\nn\\
\rd^D\kappa_l&=&\frac{\mu^{4-D}}{i\pi^{\frac{D}{2}}}\,\rd^D k_l\;,\;
P_j(\{k\},\{p\},m_j^2)=q_j^2-m_j^2+i\delta\;,
\end{eqnarray}
where the $q_j$ are linear combinations of external momenta $p_i$ and loop momenta $k_l$.
Introducing Feynman parameters leads to
\be
G=\frac{(-1)^{N_{\nu}}}{\prod_{j=1}^{N}\Gamma(\nu_j)}\Gamma(N_{\nu}-LD/2)\int
\limits_{0}^{\infty} 
\,\prod\limits_{j=1}^{N}dx_j\,\,x_j^{\nu_j-1}\,\delta(1-\sum_{l=1}^N x_l)
\frac{{\cal U}^{N_{\nu}-(L+1) D/2}}
{{\cal F}^{N_\nu-L D/2}}\,,\,N_\nu=\sum_{j=1}^N\nu_j\;,
\label{eq:F}
\ee
where ${\cal U}$ is a polynomial of degree $L$ and 
${\cal F}$ is of degree $L+1$ in the Feynman parameters. The
Lorentz invariants which can be formed from the external momenta of the diagram,
as well as propagator masses, are contained in ${\cal F}$. 
As a simple example, consider the function ${\cal F}$ for a massless one-loop box diagram:
\be
{\cal F}= -s_{12}\,x_2 x_4 -s_{23}\,x_1 x_3 -i\,\delta\;.
\label{eq:1lbox}
\ee

The functions ${\cal U}$ and ${\cal F}$  can also be constructed directly 
from the topology of the corresponding 
Feynman graph~\cite{Nakanishi,Tarasov:1996br}, 
and the implementation of this construction in \textsc{SecDec} version 2.0
is one of the new features of the program. 

For a diagram with massless propagators, 
none of the Feynman parameters occurs quadratically in 
the function ${\cal F}={\cal F}_0$ . If massive internal lines are present, 
${\cal F}$ gets an additional term 
${\cal F}(\vec x) =  {\cal F}_0(\vec x) + {\cal U}(\vec x) \sum\limits_{j=1}^{N} x_j m_j^2$. 
${\cal U}$ is a positive semi-definite function. 
A vanishing ${\cal U}$ function is related to the  UV subdivergences of the graph. 
In the region where all invariants formed from external momenta are negative, 
which we will call the {\em Euclidean region} in the following, 
${\cal F}$ is also a positive semi-definite function. 
Its vanishing does not necessarily lead to 
an IR singularity. Only if some of the invariants are zero, 
for example if some of the external momenta
are light-like, the vanishing of  ${\cal F}$  may induce an IR divergence.
Thus it depends on the {\em kinematics}
and not only on the topology (like in the UV case) 
whether a zero of ${\cal F}$ leads to a divergence or not. 
The necessary (but not sufficient) conditions for a divergence 
are given by the Landau equations~\cite{Landau:1959fi}:
\bea
&& x_j\,(q_j^2-m_j^2)=0 \quad \forall \,j \label{landau1}\\
&&\frac{\partial}{\partial k_l^\mu}\sum_j x_j\,\left(q_j^2(k,p)-m_j^2\right)=0\;.\label{landau2}
\eea
 If all kinematic invariants formed by external momenta are 
of the same sign, the necessary condition ${\cal F}=0$ for an IR divergence can only 
be fulfilled if some of the parameters $x_i$ go to zero.
These singularities can be regulated by dimensional regularisation and 
factored out of the function ${\cal F}$ using sector decomposition. 
The same holds for dimensionally regulated UV singularities contained in ${\cal U}$. 
However, after these  singularities, 
appearing as poles in  1/$\epsilon$, have been extracted using sector decomposition,
for non-Euclidean kinematics 
we are still faced with integrable singularities related to kinematic thresholds. 
How we deal with these singularities will be described in Section \ref{sec:contour}.
 
\subsection{General parameter integrals}

The program can also deal with parameter integrals which are  more 
general than the ones related to multi-loop integrals, 
for example phase space integrals involving massless particles, 
where the regions in phase space corresponding to soft and/or collinear configurations 
lead to singularities which can be extracted as poles in $1/\eps$. 
The only restrictions are that the integration domain should be the unit hypercube, 
and the singularities should be only endpoint singularities, i.e. should be located at zero 
or one. We assume that the singularities are regulated 
by non-integer powers of the integration parameters, where the non-integer part is the 
$\eps$ of dimensional regularisation or some other regulator.
The general form of the integrals is 
\be
I=\int_0^1 dx_1 \ldots \int_0^1 dx_N \prod_{i=1}^m P_i(\vec{x},\{\alpha\})^{\nu_i}\;,
\label{eq:general}
\ee
where $P_i(\vec{x},\{\alpha\})$ are polynomial functions of the parameters $x_j$,
which can also contain some symbolic constants $\{\alpha\}$. 
The user can leave the parameters $\{\alpha\}$ symbolic during the decomposition,
specifying numerical values only for the numerical integration step.
This way the decomposition and subtraction steps do not have to be redone if 
the values for the constants are changed.
The $\nu_i$ are powers of the form $\nu_i=a_i+b_i\epsilon$ 
(with  $a_i$ such that the integral is convergent).
Note that half integer powers are also possible.

\section{The \secdec{} program}

\begin{figure}[htb]
\includegraphics[width=14cm]{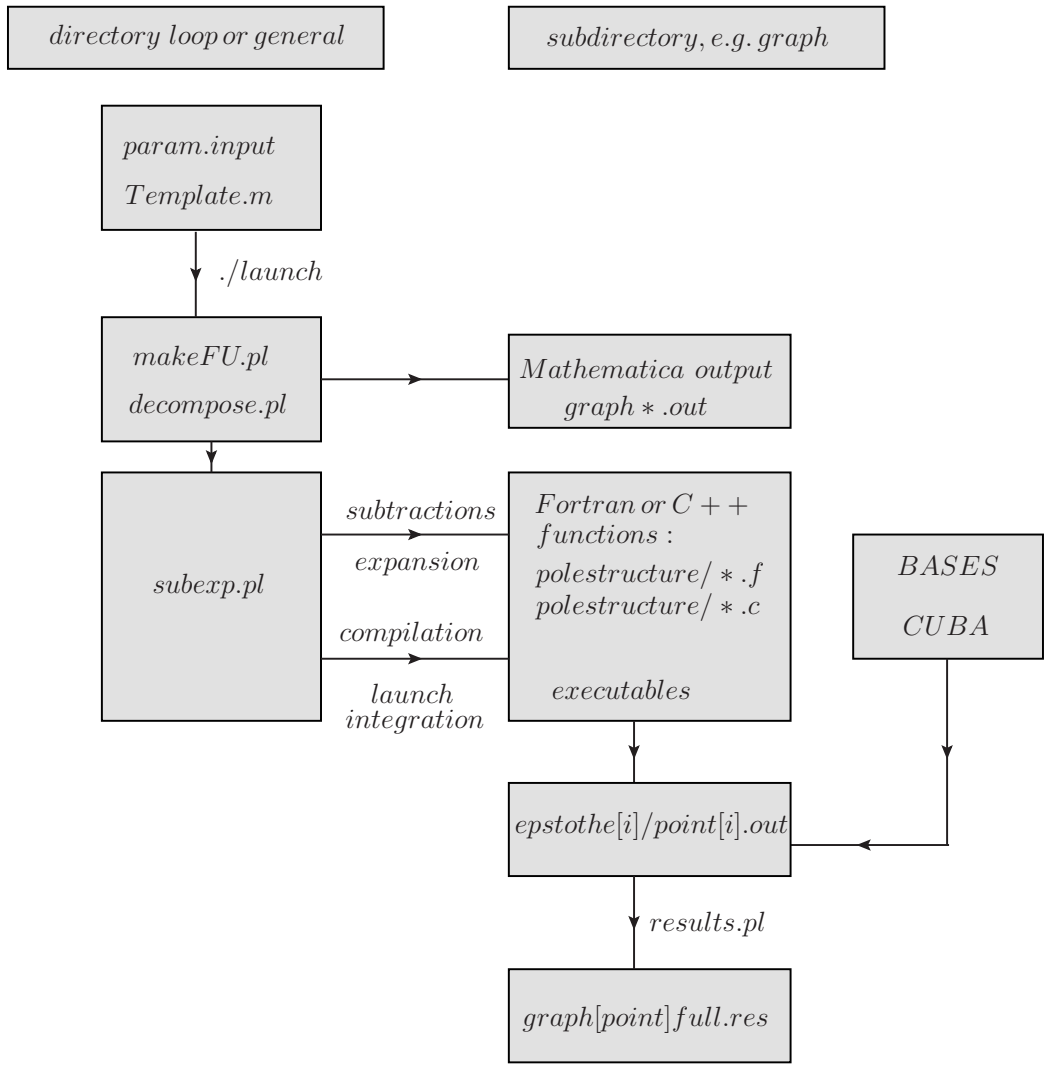}
\caption{Flowchart showing the main steps the program 
performs  to produce the result files. In each of the subdirectories 
{\tt loop} or {\tt general}, the file {\tt Template.m} can be used to define the integrand. 
The produced files are written to a subdirectory created according to the settings given 
in {\tt param.input}. By default, a subdirectory with the name of the graph or integrand is created to store the produced functions. 
This directory will contain subdirectories according to the pole structure of the integrand.}
\label{fig:flowchart}
\end{figure}

The program consists of two parts, an algebraic part and a numerical part.
The algebraic part uses code written in Mathematica~\cite{Wolfram} and does 
the decomposition into sectors, the subtraction of the 
singularities, the expansion in $\eps$ and the generation of the 
files necessary for the numerical integration. 
In the numerical part,  Fortran or C++ functions forming the coefficient 
of each term in the Laurent series in $\eps$ are integrated using the 
Monte Carlo integration programs contained in the
\textsc{Cuba} library\,\cite{Hahn:2004fe,Agrawal:2011tm}, or 
\textsc{Bases}\,\cite{Kawabata:1995th}. 
The different subtasks are handled by perl scripts.  
The flowchart of the program is shown in Fig.~\ref{fig:flowchart} for the basic 
flow of input/output streams.

\clearpage

The  directories {\tt loop} and {\tt general} have the same global structure, 
only some of the individual files are specific to loops or to more general parametric functions.
The directories contain a number of perl scripts steering the 
decomposition and the numerical integration.  The scripts use perl modules contained 
in the subdirectory {\tt perlsrc}.

The Mathematica source files 
are located in the subdirectories {\tt src/deco} (files used for the decomposition), 
{\tt src/subexp} (files used for the pole subtraction and expansion in $\eps$) and 
{\tt src/util} (miscellaneous useful functions).  
The documentation, created by {\it robodoc}\,\cite{robodoc} 
is contained in the subdirectory {\tt doc}. It contains an index to look up documentation 
of the source code in html format by loading {\tt masterindex.html} into a browser.

In order to use the program, the user only has to edit the following two files:
\begin{itemize}
\item {\tt param.input}: (text file)\\
specification of paths, type of integrand, 
order in $\eps$, output format, parameters for numerical integration, 
further options
\item {\tt Template.m}: (Mathematica syntax)
\begin{itemize}
\item for loop integrals: specification of loop momenta and propagators, 
resp. of the topology; optionally numerator, non-standard propagator powers, space-time dimensions
\item for general functions: specification of integration variables, integrand, variables to be split
\end{itemize}
\end{itemize}
The program comes with example 
input and template files in the subdiretories {\tt loop/demos} 
respectively {\tt general/demos}, described in detail in  \cite{Carter:2010hi}.

\section{Installation and usage}\label{sec:install}

\subsection{Installation}
The program can be downloaded from \\
{\tt http://projects.hepforge.org/secdec}.

Unpacking the tar archive via 
{\it  tar xzvf SecDec.tar.gz} 
will create a directory called {\tt SecDec} 
with the subdirectories as described above. Then change to the {\tt SecDec} directory
and run {\it ./install}.

Prerequisites are Mathematica, version 6 or above, perl (installed by default on 
most Unix/Linux systems), a Fortran compiler (e.g. gfortran, ifort), or a C++ 
compiler if the C++ option is used.

\subsection{Usage}

\begin{enumerate}

\item Change to the subdirectory {\tt loop} or {\tt general}, depending 
on whether you would like to calculate a loop integral or a more general parameter integral.
\item Copy the files {\tt param.input} and {\tt Template.m} to 
create your own parameter and template files  {\tt myparamfile}, {\tt mytemplatefile}.
\item Set the desired parameters in {\tt myparamfile} and specify the integrand in {\tt mytemplatefile}.
\item Execute the command {\it ./launch -p myparamfile -t mytemplatefile} 
in the shell.  \\
If you omit the option {\it -p myparamfile}, the file {\tt param.input} will be taken as default.
Likewise, if you omit the option {\it -t mytemplatefile}, 
the file {\tt Template.m} will be taken as default.
If your files {\it myparamfile, mytemplatefile} are in a different directory, say, 
{\it myworkingdir}, 
 use the option {\bf -d myworkingdir}, i.e. the full command then looks like 
 {\it ./launch -d myworkingdir -p myparamfile -t mytemplatefile}, 
 executed from the directory {\tt SecDec/loop} or
 {\tt SecDec/general}. \\

\item Collect the results. Depending on whether you have used a single machine or 
submitted the jobs to a cluster, the following actions will be performed:
 \begin{itemize}
\item If the calculations are done sequentially on a single machine, 
    the results will be collected automatically (via {\tt results.pl} called by {\tt launch}).
    The output file will be displayed with your specified text editor.

\item If the jobs have been submitted to a cluster,    
	when all jobs have finished,  execute  the command 
	{\it ./results.pl [-d myworkingdir -p myparamfile]}. 
	This will create the files containing the final results in the {\tt graph} subdirectory
	specified in the input file.

\end{itemize}


\item After the calculation and the collection of the results is completed, 
you can use the shell command {\it ./launchclean[graph]}
to remove obsolete files.

\end{enumerate}

It should be mentioned that the code starts working first on the most complicated 
pole structure, which takes longest. 
This is because in case the jobs are sent to a cluster, it is advantageous to 
first submit the jobs which are expected to take longest.

\subsection{New features}
Version 2.0 of  \secdec{} contains the following new features, 
some of which will be illustrated by examples in Section \ref{sec:examples}. 
More details are given in\,\cite{Borowka:2012yc}.
\begin{itemize}
\item Multi-scale loop integrals can be evaluated without restricting the kinematics 
to the Euclidean region.
\item The possibility to loop over ranges of parameter values is automated, 
with the option of outputting results in a format suitable for plotting.
\item The user can define functions at a symbolic level and specify them only later 
after the integrand has been transformed into a set of finite functions for each order in 
$\epsilon$. 
\item The regulator of the parameter integrals can be different from the dimensional 
regulator $\epsilon$. This is particularly useful to define e.g. measurement 
functions  at a later stage of the calculation.
\item For scalar multi-loop integrals, the integrand can be constructed 
directly from the topology of the diagram. 
The user only has to provide the labels of the vertices connected by the propagators
and the propagator masses, 
but does not have to provide the momentum flow.
\item The files for the numerical integration  of multi-scale loop 
integrals are written in C++ rather than Fortran. 
For integrations in Euclidean space, both the Fortran and the C++ versions are supported.
\item Both the algebraic and the numerical part allow  full parallelisation.
\end{itemize}

\subsection{Implementation of the contour deformation}
\label{sec:contour}

Unless the function ${\cal F}$ in equation~(\ref{eq:F}) is of definite sign for
all possible values of invariants and Feynman parameters, 
the integrand of a multi-loop integral will vanish within the integration 
region on a hypersurface given by the solutions of the Landau equations 
(\ref{landau1}),(\ref{landau2}). However, we can avoid the poles on the real axis 
by a deformation of the integration contour into the complex plane. 
As long as the deformation is in accordance with the causal $i\delta$  prescription 
of the Feynman propagators, and no poles are crossed while changing the integration
path, we can make use of Cauchy's theorem to choose an integration contour such that 
the integration is convergent. The $i\delta$ prescription for the Feynman propagators 
tells us that the contour deformation into the complex plane should be such that 
the imaginary part of ${\cal F}$ 
should always be negative. 
For real masses and Mandelstam invariants $s_{ij}$, the following Ansatz\,\cite{Soper:1999xk,Nagy:2006xy,Binoth:2005ff,Anastasiou:2007qb}
is convenient:
\bea
\label{eq:condef}
\vec{z}( \vec x) &=& \vec{x} - i\;  \vec{\tau}(\vec{x})\;,\nonumber\\
\tau_k &=& \lambda\, x_k (1-x_k)
\sum\limits_{j=1}^{N-1} \, \frac{\partial {\cal F}}{\partial x_j}  \;.
\eea
In terms of the new variables, we thus obtain
\be
\label{eq:newF}
{\cal F}(\vec{z}(\vec{x}))={\cal F}(\vec{x})
-i\,\lambda\,\sum\limits_{j=1}^{N-1} \, \left(\frac{\partial {\cal
      F}}{\partial x_j}  \right)^2 + {\cal O}(\lambda^2)\;,
\ee
such that ${\cal F}$ acquires a negative imaginary part of order
$\lambda$. 

The convergence of the numerical integration can be improved significantly 
by choosing an ``optimal" value for $\lambda$.
Values of $\lambda$ which are too small lead to contours which are too close 
to the poles on the real axis and therefore lead to bad convergence.
Too large values of $\lambda$ can modify the real part of the function 
to an unacceptable extent and could even change the sign of the
imaginary part if the terms of order $\lambda^3$ get larger than the 
terms linear in $\lambda$. This would lead to a wrong result.
Therefore we implemented a four-step procedure to optimize the value of $\lambda$, 
consisting of 
\begin{itemize}
\item ratio check: To make sure that the terms of order $\lambda^3$ in
  equation~(\ref{eq:newF})  do not spoil the sign of the imaginary part, we
  evaluate the ratio  of the terms linear and cubic in  $\lambda$ for
  a quasi-randomly chosen set of sample points to determine a maximal
allowed $\lambda=\lambda_{max}$.
\item modulus check: The imaginary part is vital at the points where
  the real part of  ${\cal F}$ is vanishing. In these regions,
  the deformation should be large enough to avoid large numerical
  fluctuations due to a highly peaked integrand. Therefore we check
  the modulus of  each subsector function ${\cal F}_i$ at a number of sample points,
and pick the fraction of the value of $\lambda_{max}$ 
maximising the function with the minimal modulus, 
i.e. the value of lambda which keeps $\mathcal F$ furthest from zero.
\item individual $\lambda(i,j)$ adjustments: If the values of
  $\frac{\partial {\cal F}_i}{\partial x_j}$  are very different in magnitude, it can be convenient to
  have an individual parameter $\lambda(i,j)$ for each subsector
  function ${\cal F}_i$ and each Feynman parameter $x_j$. 
\item sign check: After the above adjustments to $\lambda$ have been made, 
the sign of Im$(\mathcal F)$ is again checked for a number of sample
points. If the sign is ever positive, this value of $\lambda$ is disallowed.
\end{itemize}

\section{Examples and results}
\label{sec:examples}

\begin{figure}[!htbp]
\begin{center}
\includegraphics[width=5.cm]{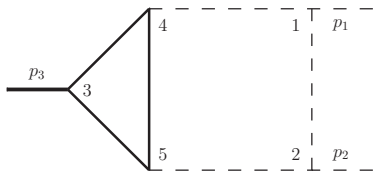}
\end{center}
\caption{Two-loop vertex graph called $P_{126}$, containing a massive triangle loop.  
Bold lines are massive, dashed lines are massless. The vertices are labeled to match the 
construction of the integrand from the topology as explained in the text.}
\label{fig:P126}
\end{figure}

In this example, we will demonstrate three of the new features 
of the \secdec{} 2.0 program: the construction of  $\mathcal F, \mathcal U$  directly from the
topology of the graph,  the evaluation of the graph in
the physical region, and how results for a whole set of 
different numerical values for the  invariants can be produced and plotted in an automated way.
We will use the two-loop diagram shown in Fig.~\ref{fig:P126} as an example. 
Numerical results for this diagram have been produced in \cite{Bonciani:2003hc,Ferroglia:2003yj}, 
analytical ones in \cite{Fleischer:1997bw,Davydychev:2003mv}.

\begin{figure}[h]
\begin{minipage}{19pc}
\includegraphics[width=15pc,angle=-90]{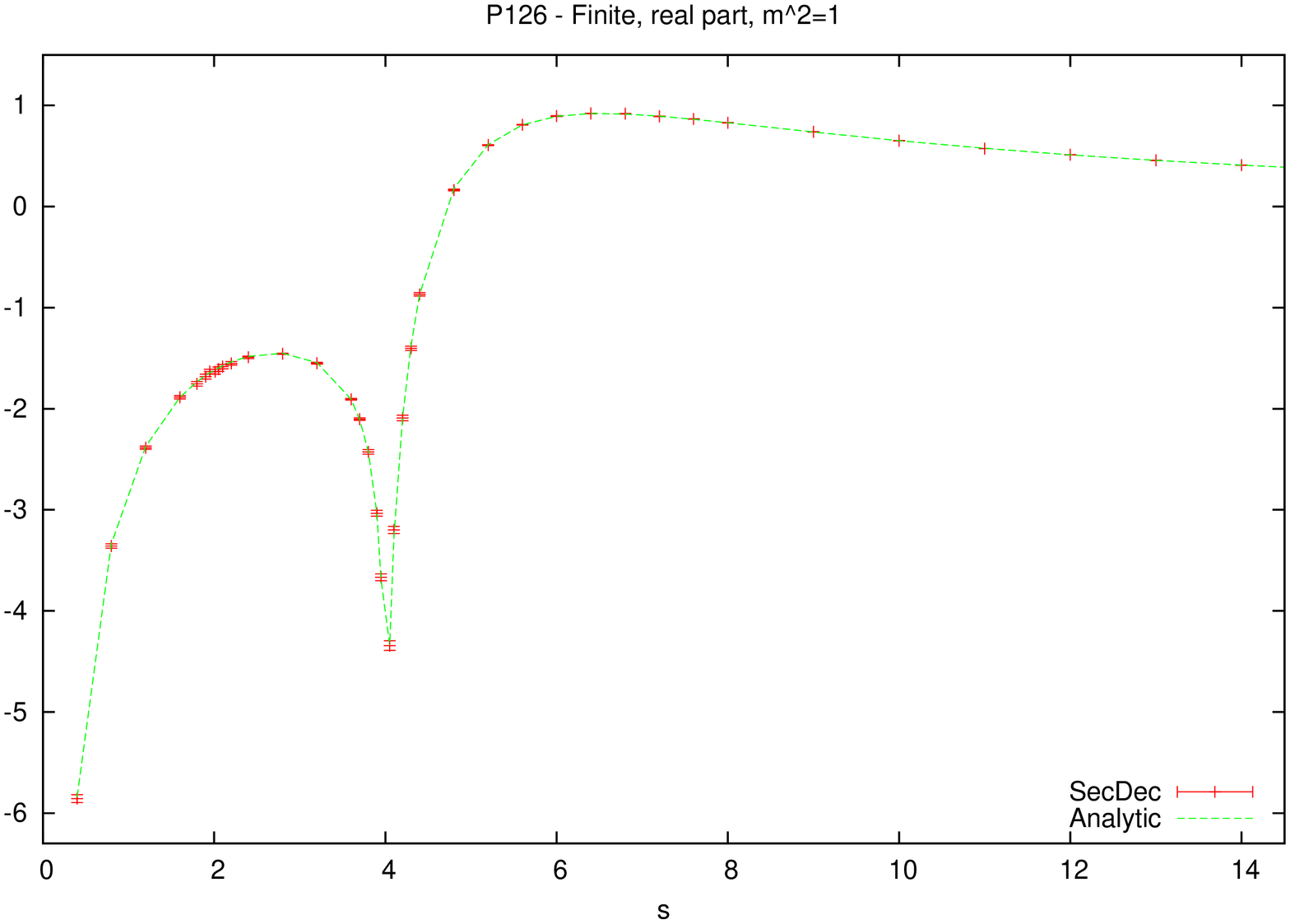}
\end{minipage}\hspace{2pc}%
\begin{minipage}{19pc}
\includegraphics[width=15pc,angle=-90]{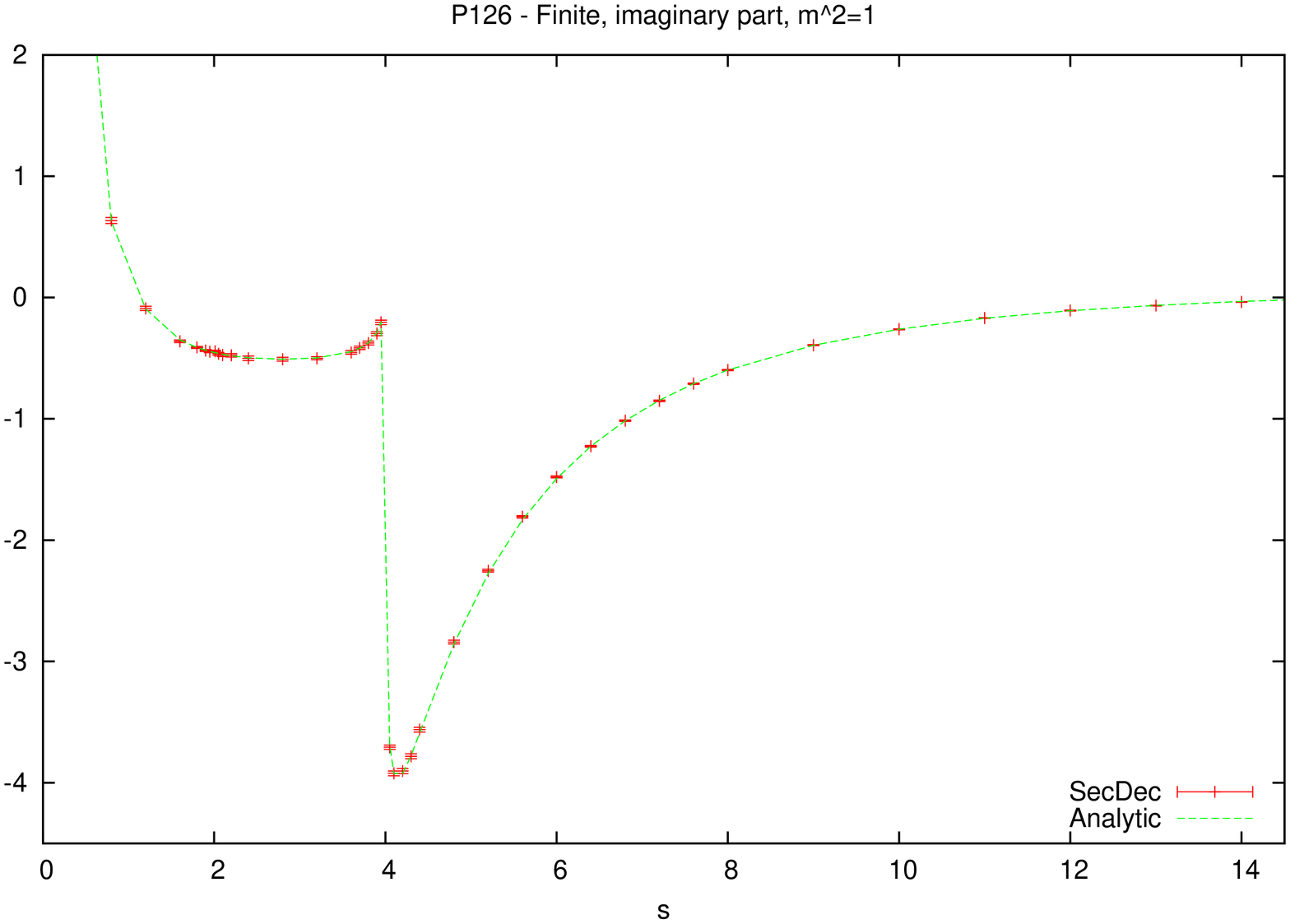}
\end{minipage} 
\caption{\label{fig:P126result}
Comparison of analytic and numerical results for the diagram $P_{126}$.}
\end{figure}

\subsection{Topology-based construction of the integrand}
The template  file {\tt templateP126.m} in the {\tt demos} subdirectory 
contains the following lines:\\
$proplist=\{\{ms[1],\{3,4\}\},\{ms[1],\{4,5\}\},\{ms[1],\{5,3\}\},\{0,\{1,2\}\},\{0,\{1,4\}\},\{0,\{2,5\}\}\};$\\
$onshell=\{ssp[1]\to 0,ssp[2]\to 0,ssp[3]\to sp[1,2]\};$\\
where each entry in $proplist$ corresponds to a propagator of the diagram; the first entry is the mass of the
propagator, and the second entry contains the labels of the two vertices which the propagator connects. 
Note that if an external momentum $p_k$ is flowing into the vertex, the vertex also must have the label $k$.
For vertices containing only internal propagators the labeling is arbitrary.
The on-shell conditions in the above example state that
$p_1^2=p_2^2=0,\,p_3^2=s$.

\subsection{Results in the physical region}
To run this example, from the {\tt loop} directory, issue the command
{\tt ./launch -d demos -p paramP126.input -t templateP126.m}.
The timings for the finite part and a relative accuracy better than 1\%, 
using \textsc{Cuba}-3.0\,\cite{Agrawal:2011tm}, are about 100 seconds
per point on an Intel(R) Core i7 CPU
at 2.67GHz, where the timings are very similar 
far from the $s=4m^2$ threshold and close to threshold. 

\subsection{Producing data files for sets of numerical values}
To loop over a set of numerical values for the invariants  $s$ and $m^2$
once the C++ files are created, issue the command\\
{\tt ./multinumerics.pl -d demos -p multiparamP126.input}. 
This will run the numerical integrations for the values of $s$ and $m^2$ specified
in the file {\tt demos/multiparamP126.input}. 
The files containing the results will be found in {\tt demos/2loop/P126}, and the
files {\tt p-2.gpdat, p-1.gpdat} and {\tt p0.gpdat} will contain the  
data files for each point, corresponding to the coefficients of 
$\eps^{-2},\eps^{-1}$ and $\eps^0$ respectively. 
These files can be used to plot the results against the analytic results using gnuplot. 
This will produce the files {\tt P126R0.ps, P126I0.ps} which will look like Fig.~\ref{fig:P126result}.


\section{Conclusions}

We have presented the program \textsc{SecDec} 2.0, which can be used to 
factorise dimensionally regulated singularities and numerically 
calculate multi-loop integrals in an automated way. 
As a new feature of the program, it now can deal with physical 
kinematics, i.e. is not restricted to the Euclidean region anymore.
A new construction of the integrand,  based entirely on topological rules, 
is also included, and the new features are demonstrated 
for the examples of a massive two-loop three-point function.
In addition, the program  can 
produce numerical results for more general parameter integrals, as they occur 
for example in phase space integrals for multi-particle production with several 
unresolved massless particles, and offers the possibility to include symbolic 
functions which can be used to define measurement functions like e.g. jet algorithms 
at a later stage. 
We are looking forward to applications of the program for the calculation 
of higher order corrections to various observables.

\section*{Acknowledgements}
J.C. was supported by the British Science and Technology Facilities
Council (STFC).
We also acknowledge the support of the Research Executive Agency (REA)
of the European Union under the Grant Agreement number
PITN-GA-2010-264564 (LHCPhenoNet).

\section*{References}

\providecommand{\newblock}{}

\end{document}